\documentclass[iop]{emulateapj}

\usepackage[T1]{fontenc}
\usepackage{natbib}
\usepackage{bm}
\usepackage[breaklinks,colorlinks,citecolor=blue]{hyperref}
\usepackage[all]{hypcap}
\usepackage{bm}
\usepackage{color}
\usepackage[dvipsnames]{xcolor}
\usepackage{setspace}
\usepackage{graphicx}
\usepackage{amsmath}

\renewcommand{\arraystretch}{0.6}

\shorttitle{}
\shortauthors{}

\begin{document}

\title{Effects of Anisotropic Viscosity on the Evolution of Active Galactic Nuclei Bubbles in Galaxy Clusters}

\author{Matthew Kingsland\altaffilmark{1},
H.-Y.\ Karen Yang\altaffilmark{1},
Christopher S.\ Reynolds\altaffilmark{3},
John A.\ Zuhone\altaffilmark{4}} 
\altaffiltext{1}{Department of Astronomy, University of Maryland, College Park, MD, USA} 
\altaffiltext{2}{Institute of Astronomy, University of Cambridge, Cambridge, UK}
\altaffiltext{3}{Harvard-Smithsonian Center for Astrophysics, Cambridge, MA, USA}
\email{Email: hsyang@astro.umd.edu}

\begin{abstract}

The interaction between jets from active galactic nuclei (AGNs) and the intracluster medium (ICM) provides key constraints on the feeding and feedback of supermassive black holes. Much understanding about AGN feedback is gained from purely hydrodynamic models; however, whether such an approximation is adequate for the magnetized, weakly collisional ICM needs to be critically examined. For example, AGN-blown bubbles in hydrodynamic simulations are easily disrupted by fluid instabilities, making it difficult to explain the coherence of observed bubbles such as the northwest ghost bubble in Perseus. In order to investigate whether magnetic tension and viscosity in realistic conditions could preserve the bubble integrity, we performed the first Braginskii-magnetohydrodynamic simulations of jet-inflated bubbles in a medium with tangled magnetic field. We find that magnetic tension alone is insufficient to prevent bubble deformation due to large velocity shear at early stage of the evolution. Although unsuppressed anisotropic viscosity in tangled magnetic field can have similar effects as isotropic viscosity, when the pressure anisotropy is bounded by microscopic plasma instabilities, the level of viscosity is substantially limited, thereby failing to prevent bubble deformation as in the inviscid case. Our results suggest that Braginskii viscosity is unlikely to be the primary mechanism for suppressing the fluid instabilities for AGN bubbles, and it remains a challenging task to reproduce smooth and coherent bubbles as observed. Because the dynamical influence and heating of the ICM critically depend on the bubble morphology, our study highlights the fundamental role of ``microphysics'' on the macroscopic properties of AGN feedback processes.

\end{abstract}



\section{Introduction}\label{intro}


Feeding and feedback of supermassive black holes (SMBH) are crucial processes determining the evolution of galaxies and galaxy clusters. Despite being the most promising mechanism for solving the ``cooling-flow problem'' in cool-core (CC) clusters \citep[][]{McNamara12}, the details of active-galactic-nucleus (AGN) feedback to the intracluster medium (ICM) remain highly debated. 


A lot of our understanding about AGN feedback is gained by purely hydrodynamic simulations, from simulations of bubble-ICM interaction \citep[e.g.,][]{Churazov01, Omma04, Guo18}, to simulations of self-regulated AGN feedback \citep[e.g.,][]{Sijacki07, Yang12a, Li14, Prasad17}. These advances have provided valuable insights into the fundamental processes of chaotic cold accretion \citep[e.g.,][]{Pizzolato05, Gaspari13} and thermalization and distribution of the jet energy \citep[e.g.,][]{Yang16b, Li17, R17, Martizzi19}. 


Despite the substantial progress, whether ideal hydrodynamic models are good representations of the ICM is still an open question. 
For instance, hydrodynamic bubbles tend to be elongated, whereas the ``fat" bubbles near the center of Perseus may be inflated by cosmic-ray (CR) dominated jets \citep[][]{Guo11, Yang19}. The morphology of the ``ghost" bubble in the northwest (NW) region of Perseus is also nontrivial to reproduce by purely hydrodynamic models due to the shorter timescales of hydrodynamic instabilities compared to the inferred age of the bubble \citep[$\sim 50-80$ Myr;][]{Dunn05}. 


To preserve the bubble coherence, some additional physical mechanisms were invoked, including magnetic field \citep[e.g.,][]{Robinson04, Ruszkowski08} and viscosity \citep[e.g.,][]{Reynolds05, Guo15}. Although bubbles in their simulations could be stabilized, these studies suffer from a few simplified assumptions, such as studying initially static cavities instead of jet-inflated bubbles, or using isotropic viscosity while the viscosity is expected to be anisotropic in the weakly collisional, magnetized ICM. \cite{Dong09} improved previous works by considering anisotropic/Braginskii viscosity along magnetic field lines; however, the survival of the bubbles depends on the simplistic field topology assumed. 


To this end, we perform three-dimensional (3D) magnetohydrodynamic (MHD) simulations of AGN jet-inflated bubbles and investigate the bubble evolution under the influence of anisotropic viscosity and tangled magnetic field. Specifically, we compare results for four cases: inviscid, isotropic viscosity, unsuppressed anisotropic viscosity, and anisotropic viscosity suppressed by plasma instabilities on microscopic scales. The last case is motivated by the recent findings that anisotropic viscosity in weakly collisional plasmas, which originates from pressure anisotropies (see \S~\ref{sec:method}), can be greatly suppressed due to firehose/mirror instabilities \citep[][]{Kunz14}.  

The structure of this Letter is as follows. In \S\ref{sec:method}, we summarize the simulation setup and describe our treatment of viscosity. In \S\ref{sec:results}, we present our main results regarding the morphology of the bubbles (\S\ref{sec:bub}) and the impact on the ICM (\S\ref{sec:icm}). We discuss the implications of the results in \S\ref{sec:discussion} and conclude our findings in \S\ref{sec:conclusion}.

\section{Methodology}
\label{sec:method}


We perform 3D MHD simulations of one pair of jet-inflated AGN bubbles in a Perseus-like cluster using FLASH \citep{Fryxell00}. The simulation setup for the initial ICM and magnetic field is identical to that in \cite{Yang16a}. The initial magnetic field is tangled with a coherence length of 25 kpc and the plasma beta ($\beta=p_{\rm th}/p_{\rm B}$) is $\sim 100$. We choose a coherence length that is greater than the typical size of AGN bubbles because this is the optimal condition for magnetic draping to occur and help prevent bubble disruption \citep{Ruszkowski08}. The injection of AGN energy in the simulations is purely kinetic, identical to the KIN case in \cite{Yang19}. The AGN injection has a total jet power of $5 \times 10^{45}$ erg s$^{-1}$ for a duration of 10 Myr, released along the $\pm z$ directions of the simulation domain. Radiative cooling is omitted because the central cooling time of Perseus is longer than the simulation duration (100 Myr). Four different assumptions about the ICM viscosity are explored: (A) inviscid, (B) unsuppressed isotropic viscosity, (C) unsuppressed anisotropy viscosity, and (D) anisotropic viscosity limited by the microscopic plasma instabilities. 



Viscosity in our simulations is included following the method of \cite{Zuhone15}. 
Specifically, our simulations solve the following Braginskii-MHD equations:

\begin{equation}
\label{eq:mass_cont}
\frac{\partial \rho}{\partial t} + \nabla \cdot (\rho \boldsymbol{v}) = 0,
\end{equation}

\begin{equation}
\label{eq:mom_cont}
\frac{\partial (\rho \boldsymbol{v})}{\partial t} + \nabla \cdot \Big (\rho \boldsymbol{v} \boldsymbol{v} - \frac{\boldsymbol{BB}}{4 \pi} + p_{\rm tot} \rm \boldsymbol{I} \Big ) = \rho \boldsymbol{g} - \nabla \cdot \boldsymbol{\Pi},
\end{equation}

\begin{equation}
\label{eq:energy_cont}
\frac{\partial E}{\partial t} + \nabla \cdot \Big [  \boldsymbol{v} (E + p_{\rm tot}) - \frac{\boldsymbol{B}(\boldsymbol{v} \cdot \boldsymbol{B})}{4 \pi} \Big ] = \rho \boldsymbol{g} \cdot \boldsymbol{v} - \nabla \cdot (\boldsymbol{\Pi} \cdot \boldsymbol{v}),
\end{equation}

\begin{equation}
\label{eq:mag_cont}
\frac{\partial B}{\partial t} + \nabla \cdot (\boldsymbol{vB} - \boldsymbol{Bv}) = 0,
\end{equation}
where $p_{\rm tot}=p+B^2/(8\pi)$ is the total pressure, and all other variables follow their usual definitions. 
The viscosity tensor for the isotropic case is defined as \citep{Spitzer62}
\begin{equation}
\label{eq:iso_pi}
\boldsymbol{\Pi}_{\rm iso} = - \mu \nabla \boldsymbol{v}.
\end{equation}
In the ICM, in which the gyro-radii of particles are much smaller than the Coulomb mean free path, the viscosity should be anisotropic, and the viscosity stress tensor can be expressed as \citep{Braginskii65}
\begin{equation}
\boldsymbol{\Pi}_{\rm aniso} = - 3 \mu \Big (\boldsymbol{b} \boldsymbol{b} - \frac{1}{3} \boldsymbol{I}  \Big ) \Big (\boldsymbol{b} \boldsymbol{b} - \frac{1}{3} \boldsymbol{I}  \Big ) : \nabla \boldsymbol{v},
\label{eq:pi}
\end{equation}
where $\mu = 2.2 \times 10^{-15} \ T^{5/2} / \ln \Lambda \ \rm g \ cm^{-1} \ s^{-1}$ is the dynamic viscosity coefficient ($\rm ln \ \Lambda = 30$), and $\boldsymbol{b}$ is the magnetic field unit vector. For all viscous simulations, a ceiling is applied for the kinematic viscosity coefficient ($\nu = \mu/\rho$) of $10^{30}$ cm$^2$ s$^{-1}$. This is to prevent $\mu$ from becoming unusually large within the bubbles due to high temperatures resulted from purely kinetic jets.\footnote{Previous simulations of \cite{Reynolds05} and \cite{Guo15} also used a constant $\mu$ to mitigate this effect.}


In Braginskii-MHD, the viscosity originates from the pressure anisotropy that arises due to conservation of the first and second adiabatic invariants of particles on timescales that are much greater than the inverse of the ion gyrofrequency \citep{Chew56}. Under the condition that the pressure anisotropy is balanced by its relaxation via ion-ion collisions \citep{Schekochihin05}, one can show that 
\begin{equation}
p_\perp - p_\parallel = 0.96 \frac{p_{\rm i}}{\nu_{\rm ii}} \frac{d}{dt} \ln \frac{B^3}{\rho^2} = 3 \mu \left({\bm b}{\bm b} - \frac{1}{3} \mathsf{I} \right) : \nabla {\bm v},
\label{eq:paniso}
\end{equation}
where $p_{\perp}$ and $p_{\parallel}$ are the pressure perpendicular and parallel to the magnetic field line, respectively, $p_{\rm i}$ is the ion thermal pressure, and $\nu_{\rm ii}$ is the ion-ion collisional frequency. The total thermal pressure satisfies $p=(2/3)p_\perp + (1/3)p_\parallel$. Given Eq.\ \ref{eq:paniso}, the resulting viscous stress tensor could be written in a form identical to Eq.\ \ref{eq:pi} (see Section 3.1 of \cite{Zuhone16} for a brief derivation). When the pressure anisotropy violates the inequalities
\begin{equation}
\label{eq:pbound}
- \frac{2}{\beta} \ \lesssim \ \Delta_p \equiv \frac{p_{\perp} - \ p_{\parallel}}{p} \  \lesssim \frac{1}{\beta},
\end{equation}
fast-growing firehose (which occurs when $\Delta_p < -2/\beta$) and mirror (when $\Delta_p > 1/\beta$) instabilities are triggered and the pressure anisotropies should be kept within the marginal-stability thresholds \citep{Schekochihin05, Kunz14}. To account for this effect, in case D we apply bounds to the pressure anisotropies (thus limiting viscosity) according to Eq.\ \ref{eq:pbound}. 

Note that the above means that the Braginskii-MHD equations (as used in case C and the previous study of \cite{Dong09}) become ill-posed when Eq.\ \ref{eq:pbound} is violated. Without the microscopic effects being taken into account, the fastest-growing modes of the instabilities formally occur at infinitely small scales, which are essentially the grid scale where the microinstabilities may be unresolved. Later we will see that, indeed, while case C is able to generate some modes of the firehose fluctuations, the mirror instability is not captured and hence the positive pressure anisotropy could go beyond the stability criterion, substantially overestimating the level of viscosity. Although this case is rather unphysical, we include it in this work in order to aid the interpretation of our results and to make a direct comparison with the previous work of \cite{Dong09}.


\section{Results}
\label{sec:results}
 
\subsection{Coherence of AGN bubbles} 
\label{sec:bub}

\begin{figure*}[tbp]
\begin{center}
\includegraphics[scale=0.6]{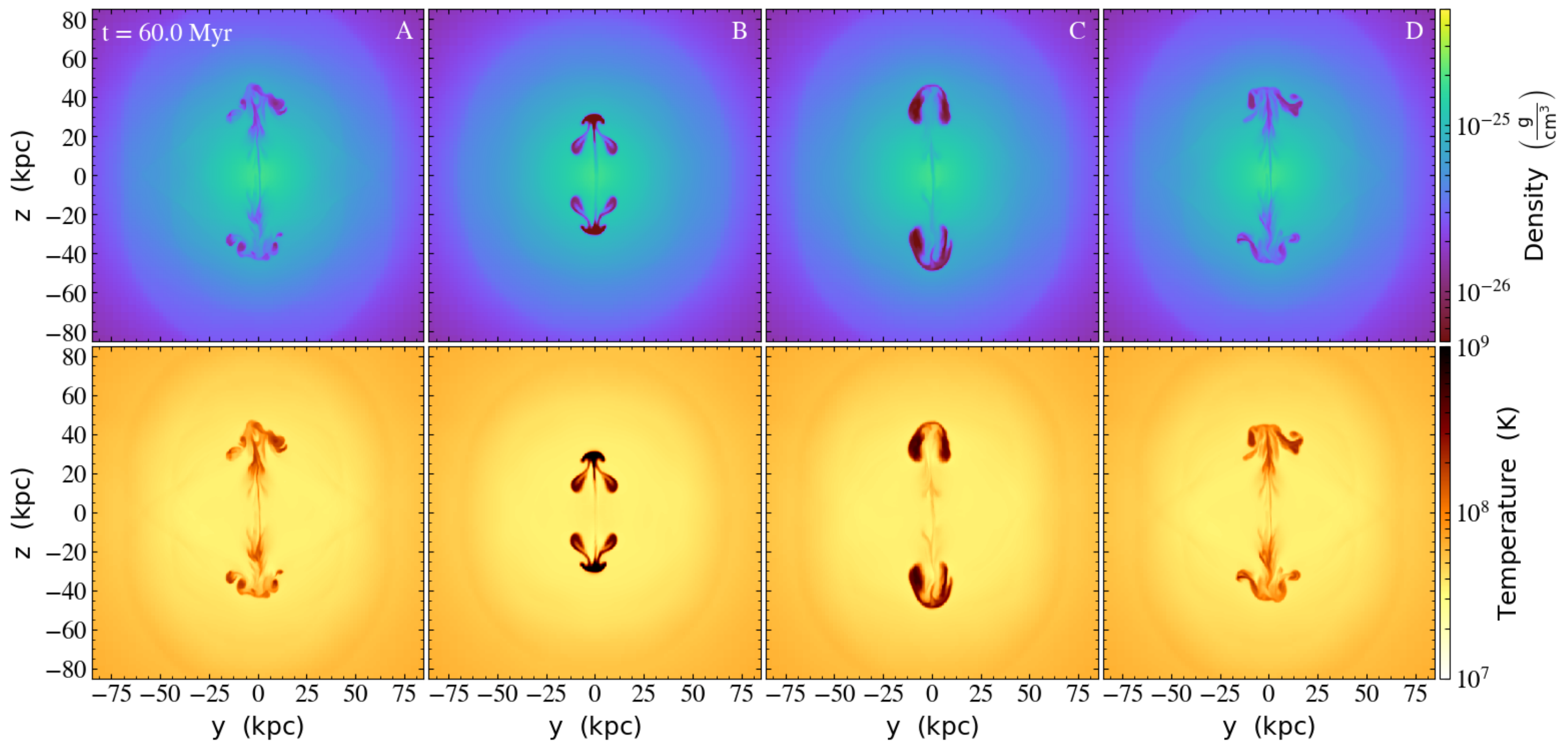} 
\caption{Slices of gas density (top) and temperature (bottom) at $t=60$ Myr for cases A (inviscid), B (unsuppressed isotropic viscosity), C (unsuppressed anisotropic viscosity), and D (anisotropic viscosity bounded by microinstabilities).}
\label{fig:denstemp}
\end{center}
\end{figure*}



Figure \ref{fig:denstemp} shows slices of gas density and temperature at $t=60$ Myr for cases A-D. In all four simulations, the energy injection from the AGN inflates bubbles that are characterized by low densities and high temperatures. One immediately notices the different bubble morphology among the four simulations with different treatments of viscosity. For the inviscid simulation (A), the bubble shapes are irregular and the surfaces are rippled due to Rayleigh-Taylor (RT) and Kelvin-Helmholtz (KH) instabilities as the low-density bubbles move through the dense ICM core with a velocity shear. As a result, the bubbles are gradually disrupted and mixed with the ambient ICM. The timescales for the growth of RT and KH instabilities at the bubble surface evaluated at $t=12$ Myr are $\sim 20$ and 6.7 Myr, respectively (for density contrast $\eta \sim 0.1$ and shear velocity $\Delta v \sim 3000$ km s$^{-1}$). Note that in order for magnetic tension to suppress the KH instabilities, $\Delta v$ has to be smaller than the rms Alfv\'en speed in the two media \citep{Chandrasekhar81}. Given the large shear velocity at early times, magnetic tension is unable to preserve the smooth surface of bubbles self-consistently inflated by AGN jets.

\begin{figure*}[tbp]
\begin{center}
\includegraphics[scale=0.6]{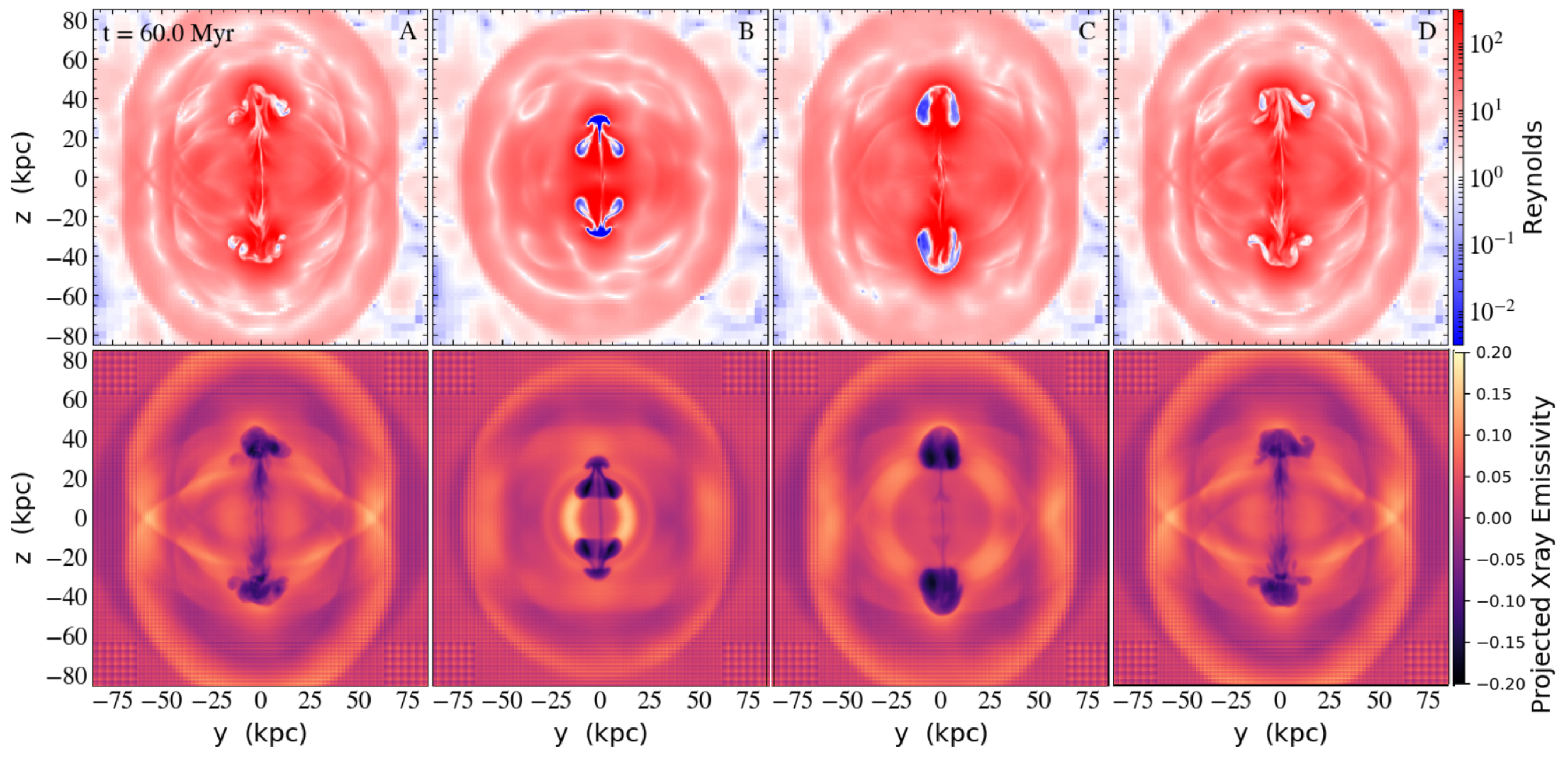}
\caption{Maps of the Reynolds number (top) and projected X-ray emissivity (bottom; fractional variation from a radially averaged projected emissivity profile) at $t=60$ Myr for cases A-D.}
\label{fig:xrayRe}
\end{center}
\end{figure*}

For the simulations with unsuppressed viscosity, either isotropic (B) or anistropic (C), the morphology of the bubbles is distinct from the inviscid case. Due to the suppression of fluid instabilities by viscosity, the bubble surface is much more smooth, and mixing is greatly inhibited (evident from the bubble-ICM density and temperature contrasts). The importance of viscosity in cases B and C can be seen from Figure \ref{fig:xrayRe}, which shows that the Reynolds numbers ($Re \equiv UL/\nu$) are $\lesssim 1$ for the bubble interior. As a result, the bubbles look much more coherent in the mock X-ray image (bottom row of Figure \ref{fig:xrayRe}) in cases B and C, in contrast to the more patchy bubbles with rippled surfaces in case A. Our result for the isotropic viscosity case confirms previous studies \citep{Reynolds05, Guo15}. In contrast to \cite{Dong09}, who showed that the coherence of bubbles depends on magnetic field topology, we show that unsuppressed anisotropic viscosity in tangled magnetic field can suppress the fluid instabilities and prevent the bubbles from disruption. Although anisotropic viscosity only inhibits the instabilities along field lines, the randomness of the tangled field helps to stabilize the bubbles in multiple directions. Therefore, its effect is very similar to the isotropic case, though the effective isotropic viscosity is somewhat suppressed with respect to the full Spitzer value due to the field geometry \citep[analogous to the factor of $\sim 1/5-1/3$ suppression of thermal conductivity;][]{Narayan01}. This is also consistent with the results of \citet{Zuhone15}, who found that the effect on suppressing the KH instabilities for cold fronts of an isotropic viscosity $\sim$1/10 of the Spitzer value was similar to Braginskii viscosity.


\begin{figure*}[tbp]
\begin{center}
\includegraphics[scale=0.6]{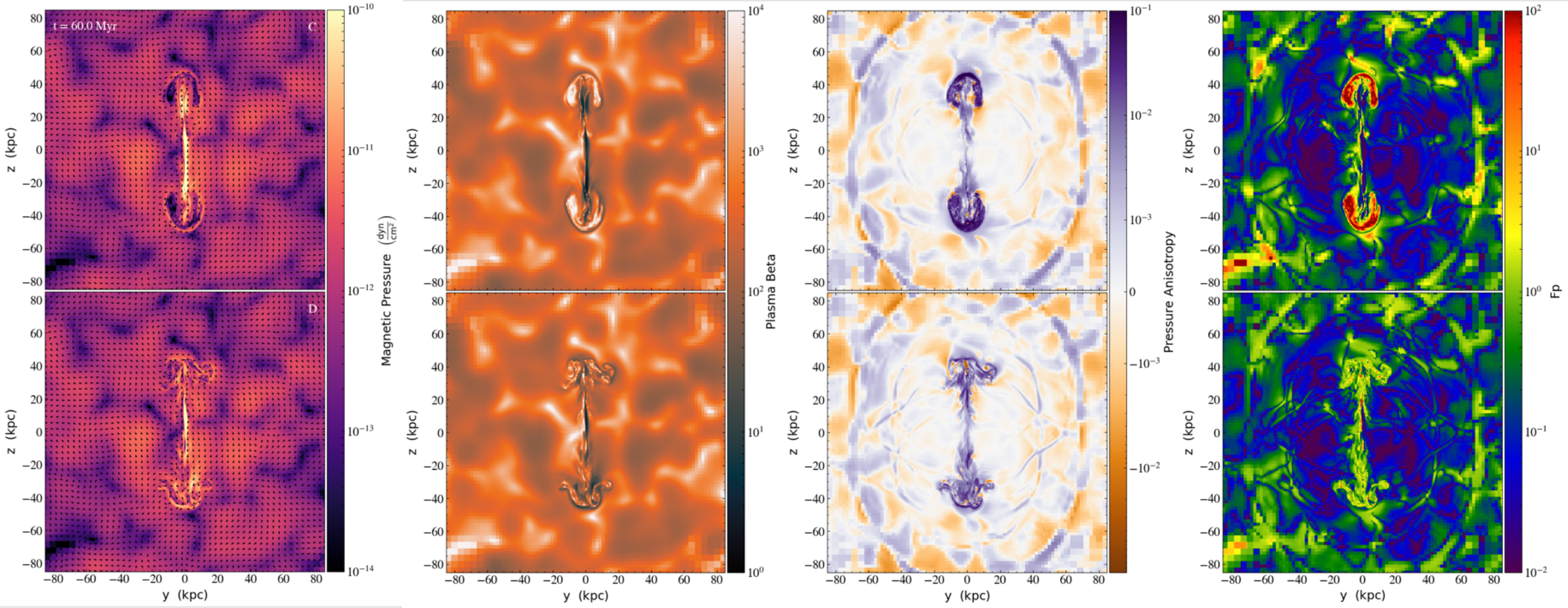} 
\caption{Columns from left to right show slices of magnetic pressure overplotted with magnetic field vectors, plasma beta, pressure anisotropy ($\Delta_{\rm p}$), and departure from marginal stability ($f_{\rm p}\equiv \beta \Delta_{\rm p}$) for cases C (top) and D (bottom).}
\label{fig:mag}
\end{center}
\end{figure*}

An even more interesting and somewhat surprising result is that, when we add the effects of microinstabilities that limit the pressure anisotropies (case D), the result is completely flipped -- anisotropic viscosity is no longer able to preserve the bubble coherence, 
as if there were no viscosity. 
We can understand this result by looking at the distributions of magnetic field and pressure anisotropies comparing cases C and D (Figure \ref{fig:mag}). Driven by converging/diverging field lines in the background tangled field, both simulations exhibit pressure anisotropies on the order of $10^{-3}$ in the ambient medium. The pressure anisotropies are most significant at the shocks and within the bubbles, due to enhanced temperatures at these locations. Without the microinstabilities, the pressure anisotropy in the bubble interior could reach $\sim 0.1$ owing to significant compressive motions by the jets. Note that the pressure anisotropies are predominantly positive because Braginskii-MHD simulations without the microinstability limiter are able to capture the firehose instability (which regulates the negative pressure anisotropies) but not the mirror instability \citep[see discussions in, e.g.,][]{Kunz12}. 

On the other hand, the bubble interior is also where the magnetic field pressure is lower due to the adiabatic expansion of the bubbles. Therefore, when the bounds for pressure anisotropies are applied, the enhanced plasma beta within the bubbles ($\beta \sim 10^4$) dramatically limits the permitted range of pressure anisotropies and thus the level of viscosity, to the degree that the bubbles are deformed by fluid instabilities just as in the inviscid simulation. In contrast to the unlimited pressure anisotropies in case C that could go $\sim$ 10-100 times beyond the marginal-stability threshold within the bubbles, for case D we find $f_{\rm p} \equiv \beta \Delta_{\rm p} \sim 1$. In other words, the microinstabilities effectively provide a factor of $\sim 10-100$ suppression of the viscosity, thereby paralyzing its ability to suppress the fluid instabilities.

\subsection{Impacts on the ICM}
\label{sec:icm}


\begin{figure*}[tbp]
\begin{center}
\includegraphics[scale=0.65]{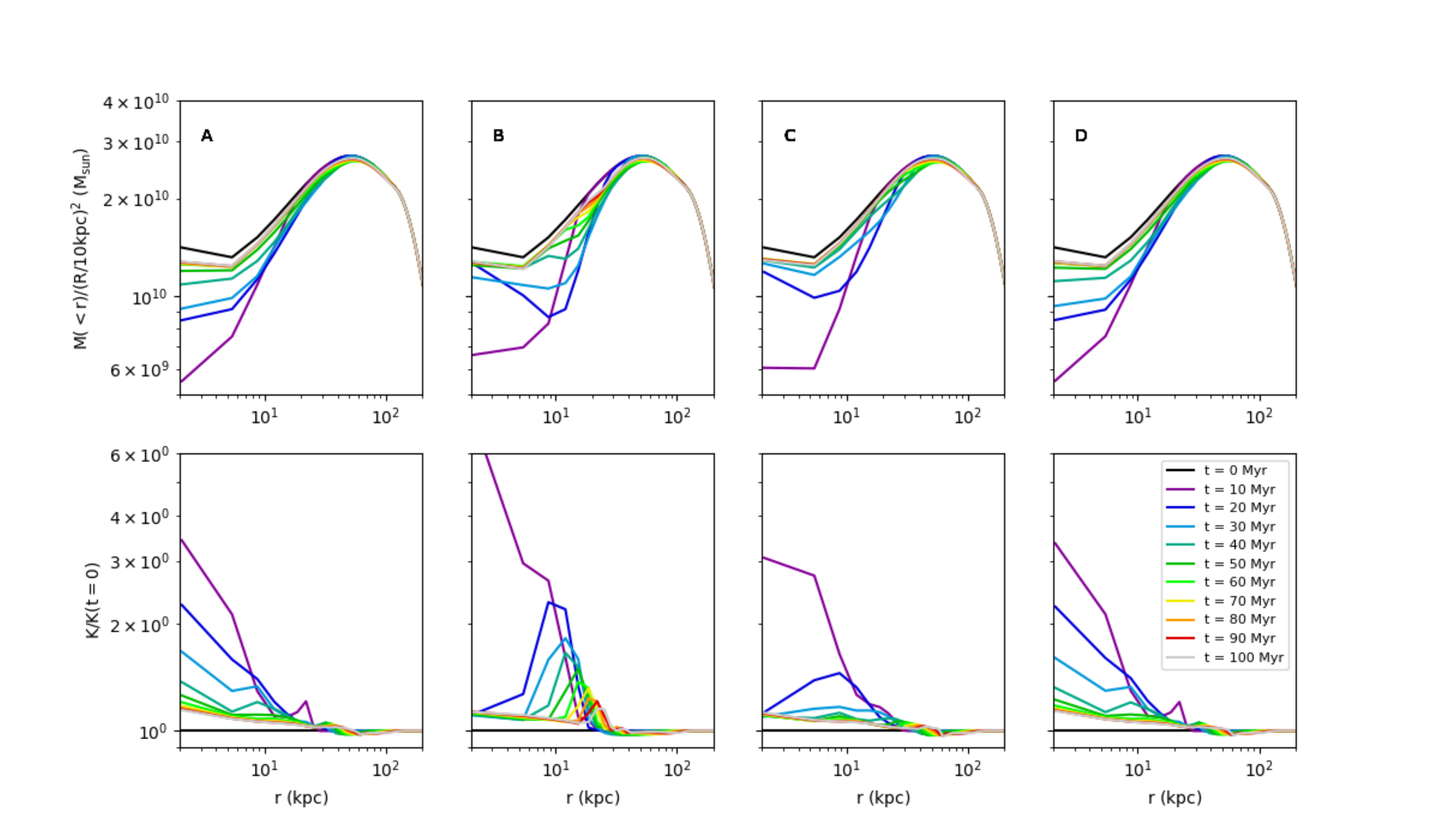} 
\caption{Evolution of the enclosed mass profiles (top; normalized by radius squared to show the variation more clearly) and entropy profiles relative to the initial state (bottom) for cases A-D.}
\label{fig:prf}
\end{center}
\end{figure*}

Figure \ref{fig:prf} shows that evolution of radial profiles of the enclosed mass and the change of gas entropy ($K \equiv T/n^{2/3}$), which traces ICM uplifting and locations of heating by the bubbles, respectively. One can see that, except for the initial transient right after the jet injection, the results could be divided into two groups: simulations in which the bubbles are deformed (A and D), and simulations in which the bubbles maintain their integrity (B and C). For the former group, the trailing part of the bubbles tends to push the ICM outward at all radii within $\sim 50$ kpc, whereas the more coherent bubbles in the latter group are more capable of uplifting the medium immediately surrounding the bubbles. For the former group, the heating primarily occurs in the wakes of the bubbles. This is where significant turbulent mixing takes place and the heating to the ICM is done by direct mixing with the ultra-hot bubbles \citep{Yang16b}. By contrast, for the latter group, the ICM is heated the most in regions surrounding the bubbles owing to both direct mixing and viscous heating. In addition, the heat is deposited further away from the cluster center as the bubbles gradually move outward and are disrupted on longer timescales.    


\begin{figure}[tbp]
\begin{center}
\includegraphics[scale=0.5]{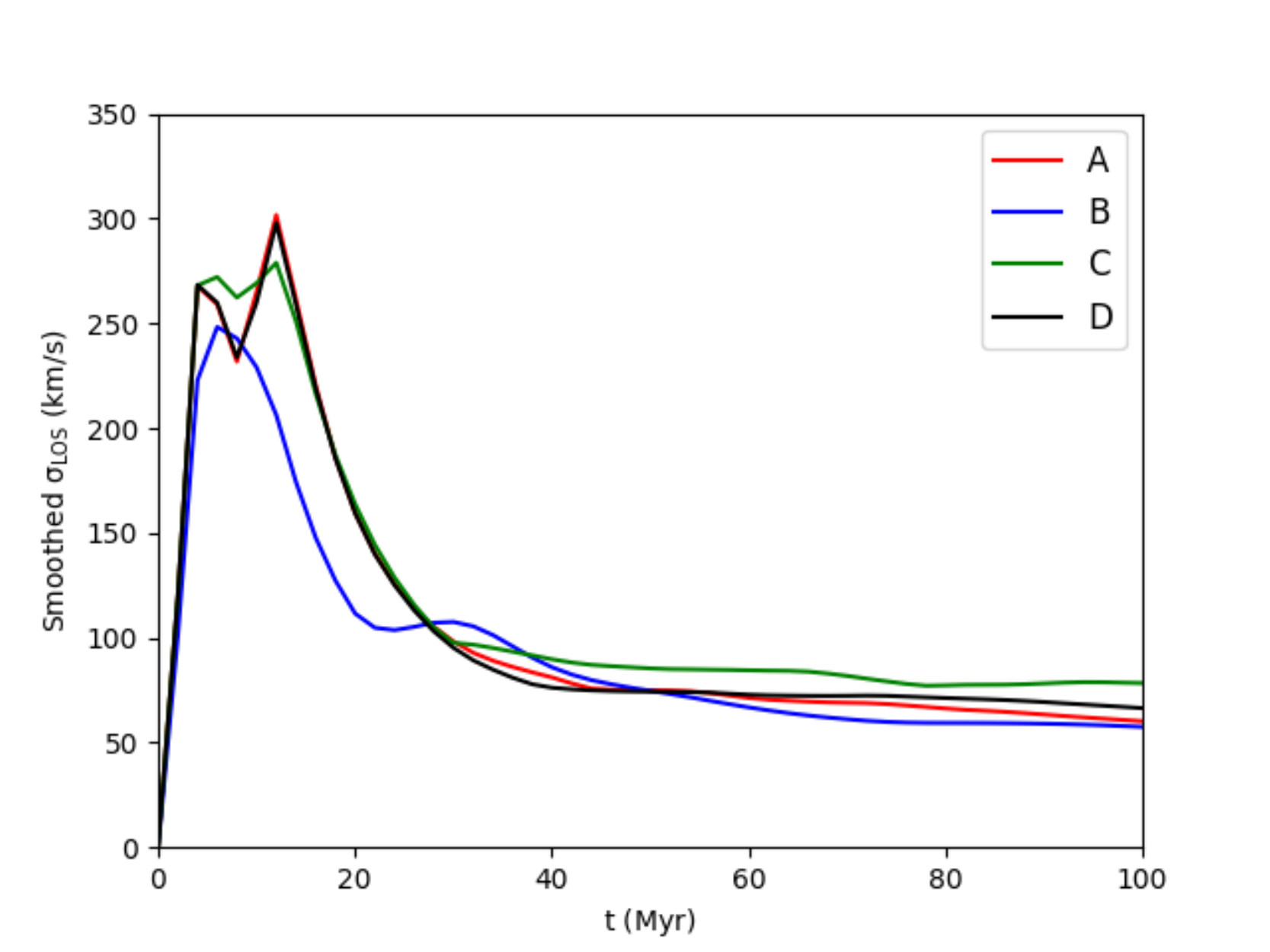} 
\caption{Time evolution of the maximum LOS velocity dispersion smoothed with {\it Hitomi} resolution for cases A-D.}
\label{fig:sigma}
\end{center}
\end{figure}

To compare the influence on the ICM kinematics by different treatments of viscosity, we compute the line-of-sight (LOS) velocity dispersion ($\sigma_{\rm LOS} \equiv (\langle v_l^2 \rangle - \langle v_l \rangle^2)^{1/2}$, where brackets represent emission-weighted averages and $v_l$ is the velocity component along the LOS, which is assumed to be the $x$-axis here) smoothed with {\it Hitomi} resolution. Figure \ref{fig:sigma} shows the maximum value across the generated $\sigma_{\rm LOS}$ map at each epoch for all cases. In general, we do not find significant differences among different simulations regarding the overall evolution of $\sigma_{\rm LOS}$: all peaks at $\sim 250-300$ km s$^{-1}$ within the first $\sim 10-20$ Myr, and decreases to $\sim 50-90$ km s$^{-1}$ after 40 Myr. The LOS velocity dispersion is relatively insensitive to viscosity because the dispersion is dominated by fluid motions on larger scales than the viscous scale \citep{Zuhone18}. 
Except for the first 20 Myr, these values are all smaller than the measured value by {\it Hitomi} of $\sim 200$ km s$^{-1}$ in the NW region or $\sim 100$ km s$^{-1}$ for other regions \citep{Hitomi18}. Note, however, that our simulations only considered a single AGN outburst. More realistic simulations of self-regulated AGN feedback will be required to determine whether any of these cases would generate too small velocity dispersions that violate observational constraints. 



\section{Discussion}
\label{sec:discussion}


Our results suggest that it remains challenging to produce smooth and coherent bubbles by momentum-driven AGN jets. While magnetic tension could suppress fluid instabilities for initially static bubbles \citep{Robinson04, Ruszkowski08}, it is more difficult to stabilize jet-inflated bubbles. While full Braginskii anisotropic viscosity in a tangled magnetic field could mimic isotropic viscosity and prevent bubble disruption, when the pressure anisotropies are bounded by microinstabilities, the level of viscosity is dramatically reduced and the bubbles are deformed just as in the inviscid simulation. Therefore, our work suggests that microinstability-limited Braginskii viscosity is unlikely to be the primary mechanism for suppressing fluid instabilities for AGN bubbles.  


In order to explain the coherent structure and smooth surface of observed bubbles, other mechanisms may be required. For instance, \cite{Scannapieco08} showed that by including a subgrid model of turbulence, bubbles with smoother surfaces can be produced due to the cancellation of small-scale modes of the fluid instabilities. Note, though, that their model of subgrid turbulence does not account for the KH instability, which is the dominant instability for bubbles inflated by momentum-driven jets. Moreover, the subgrid turbulence model employed is based on ideal hydrodynamics, and whether or not ideal hydrodynamic/MHD models are good approximations for the ICM remains an open question. Indeed, recent studies have shown that there exist fundamental differences between MHD and Braginskii-MHD turbulence \citep[][]{Squire19}. 
Clearly, further studies are demanded to understand the rich microphysics of the ICM plasma and how it impacts AGN feeding and feedback on large scales.


Though not the focus of our current study, here we briefly comment on the level of ICM viscosity in the regions excluding the bubbles. First of all, we find that the pressure anisotropies driven by shocks and sound waves produced by the AGN outburst are around the marginal-stability threshold (rightmost column of Figure \ref{fig:mag}), suggesting that the microinstabilities should have minimal effects in these regions and the level of viscosity could be close to the full Braginskii value. Interestingly, this is consistent with constraints on the parallel viscosity obtained by observations of cold fronts \citep{Zuhone15} and, assuming suppression due to fully tangled magnetic field, moderate suppression factors for the effective isotropic viscosity ($f \lesssim 5\% - 20\%$) inferred from observations of sloshing cold fronts and ram-pressure stripping tails of cluster galaxies \citep[e.g.,][]{Su17, Wang18}. 
Such a non-negligible level of viscosity would imply that viscous dissipation of sound waves could still be an important source of AGN heating \citep{Fabian03, Zweibel17b, Bambic19}.

Another implication of our work is that the viscosity of the ICM is likely highly spatially variable. While the level of viscosity is expected to be enhanced in regions of high temperatures, it can be significantly limited by the microinstabilities in regions with strong magnetic-field compression/rarefactions. This stresses the importance of Braginskii-MHD simulations with the microinstability limiter in studies of ICM transport processes, including cold-front and ram-pressure stripping simulations.




There are a few limitations of our current work. First, our results only apply to kinetic-energy-dominated AGN jets. In reality, the composition of the jets and bubbles is largely unknown \citep{Dunn04}. It remains to be seen whether viscosity is required to preserve bubbles inflated by magnetically dominated jets \citep[e.g.,][]{Li06, ONeill10}, CR dominated jets \citep[e.g.,][]{Guo11, R17, Yang19}, or internally subsonic jets in general \citep{Guo16}. In particular, magnetically dominated jets or other mechanisms such as turbulent amplification \citep[e.g.,][]{Yang13} could act to preserve/replenish the magnetic energy within the bubbles, which may yield a lower plasma beta within the bubbles and potentially alleviate the constraint on viscosity. 
In addition, our simulations have neglected pre-existing turbulence in the ICM, which could break one bubble into multiple segments and complicate observational identification of radio bubbles \citep{Heinz06}. 
This effect should be taken into account when a more detailed comparison between simulated bubbles and observed cavities is performed in the future.    


\section{Conclusions}
\label{sec:conclusion}

We performed the first 3D Braginskii-MHD simulations of momentum-driven AGN jets to investigate the effects of anisotropic viscosity on the evolution of bubbles that are self-consistently inflated by the jets and evolve in realistic, tangled magnetic field. We varied four different cases of viscosity (inviscid, unsuppressed isotropic viscosity, unbounded anisotropic viscosity, anisotropic viscosity bounded by microinstabilities) to study the morphology of the bubbles as well as its resulting impacts on the ICM. Our conclusions are as follows.

1.\ When jet inflation of the bubbles is self-consistently modeled, even magnetic field with coherence lengths greater than the bubble size cannot prevent the deformation of the bubbles.  

2.\ Unsuppressed anisotropic viscosity along tangled magnetic field lines can have similar effects as isotropic viscosity and is capable of preventing the bubbles from disruption. However, the level of viscosity in the vicinity of the bubbles in this case is overestimated by a factor of $\sim 10-100$ compared to the case with bounded pressure anisotropies. 

3.\ Adding bounding microinstabilities to the pressure anisotropy of the system drastically changes the outcome of the bubble evolution. The viscosity within the bubbles is so significantly suppressed by the microinstabilities that it can no longer prevent the bubbles from deformation, resembling the inviscid case. Note that the constraints on viscosity depend on the plasma $\beta$ within the bubbles and thus could potentially be alleviated by mechanisms that could enhance the bubble magnetic field.

4.\ The LOS velocity dispersions computed from the current simulations saturate at $\sim 50-90$ km s$^{-1}$ for all cases, which is smaller than the {\it Hitomi} measurement for the NW region. Future simulations of self-regulated feedback are required to determine whether any of these models can be ruled out by observational data from {\it Hitomi, XRISM, Athena} and {\it Lynx}.

5.\ The ability to uplift the ICM and the locations of heating by the bubbles critically depend on whether or not the bubbles are preserved. In other words, obtaining an accurate prescription for the bubble-ICM interaction remains a key question for determining the dynamical impact and heating of the ICM by AGN jets. 

Our simulations suggest that it remains a challenge to produce smooth and coherent bubbles as the NW ghost cavity in Perseus by momentum-driven AGN jets; mechanisms other than Braginskii viscosity appear to be required to suppress the fluid instabilities at bubble surfaces. Detailed comparisons between Braginskii-MHD simulations and high spatial and spectral resolution X-ray observations of AGN bubbles, ICM turbulence, cold fronts, and ram-pressure stripping tails will provide crucial constraints on the transport coefficients of the ICM. Our results also highlight the dramatic influence of the ``microphysics" on the macroscopic properties of AGN bubble evolution. Accurate modeling of the ICM plasma is thus fundamental for constructing a robust model for AGN feeding and feedback. 


\section*{\bf \scriptsize Acknowledgements}
H.Y.K.Y.\ acknowledges support from NASA ATP (NNX17AK70G) and NSF (AST 1713722). J.A.Z.\ acknowledges support from NASA contract NAS8-03060 with the {\it Chandra} X-ray center. The simulations were performed on {\tt Pleiades} at NASA and {\tt Deepthought2} at University of Maryland. FLASH was developed by the FLASH center at University of Chicago. Data analysis was conducted with the $yt$ visualization software \citep{yt}.

\bibliographystyle{biblio}
\bibliography{agn}

\end{document}